# EVIDENCE FOR ELECTROWEAK CORRECTIONS IN THE STANDARD MODEL*

ALBERTO SIRLIN
*Department of Physics, New York University, 4 Washington Place*
*New York, NY 10003, USA*
E-mail: sirlin@mafalda.physics.nyu.edu

ABSTRACT

The phenomenological evidence for electroweak corrections in the Standard Model, both at very low energies and the $Z^0$ scale, is discussed. In particular, we review a simple but sharp argument for the presence of Electroweak Bosonic Corrections.

## 1. Universality of the Weak Interactions.

Historically, the first important application involving large radiative corrections to allowed weak-interaction processes is the analysis of Universality of the Weak Interactions.[1] In modern language, the test of universality reduces to the question of whether or not the CKM matrix is unitary, a fundamental tenet of the Standard Model (SM). The most precise test involves the relation

$$|V_{ud}|^2 + |V_{us}|^2 + |V_{ub}|^2 = 1 \ . \tag{1}$$

The term $|V_{ud}|^2$ is obtained from the ratio of the decay propabilities of the eight accurately measured Fermi transitions (a well-known example is $^{14}O \to {}^{14}N + e^+ + \nu$) and $\mu$ decay, while $V_{us}$ is extracted from $K_{\ell 3}$ and hyperon decays. ($|V_{ub}|^2$ plays an essentially negligible role at present). If only the very large Fermi Coulomb corrections are included, the test does not work : the l.h.s. of Eq.(1) is found to be $\approx 1.04$, with uncertainties of $\mathcal{O}(0.1\%)$. Thus, the SM is not tenable under such a simplified analysis and it is necessary to evaluate the additional $\mathcal{O}(\alpha)$ corrections. There is however, a basic theoretical difficulty : in a rigorous analysis, one cannot simply use elementary Feynman diagrams because $\beta$-decay involves complex hadronic systems at very small momentum transfers. Instead, it is possible to express the radiative corrections in terms of current correlation functions, i.e. Fourier trasforms of matrix elements of time-ordered products of current operators, and make use of their associated Ward identities and short distance expansions.[2] Remarkably, the calculation can be carried out to good accuracy if one assumes that current conservation is softly broken, i.e. by mass terms, and that the strong interactions are asymptotically free

*To appear in the proceedings of the Ringberg Workshop on "Perspectives for electroweak interactions in $e^+e^-$ collisions" hosted by the Max Planck Institut, at the Ringberg Castle, Münich, February 5-8, 1995. Edited by B. Kniehl.

(as in QCD). In the local Fermi theory of weak interactions, the $\mathcal{O}(\alpha)$ corrections to the ratio is divergent, while the SM, being a renormalizable theory, provides a finite answer. Furthermore, for reasons that are not well understood, in versions of the $\mathrm{SU(2)_L \times U(1)}$ theory where the Higgs scalars trasform as singlets and doublets, so that $\cos^2\theta_W = M_W^2/M_Z^2$ is a natural relation, the answer is the same as in the local theory with the cutoff $\Lambda$ replaced by $M_Z$. (This assumes that the couplings of the Higgs scalars to leptons and light quarks are very small in analogy with the minimal version of the theory). Schematically, one obtains for the $\beta$-decay probability[2,3]

$$P = P^0 \left\{ 1 + \frac{3\alpha}{2\pi} \left[ \ln\left(\frac{M_Z}{2E_m}\right) + 2\bar{Q}\ln\left(\frac{M_Z}{M}\right) + ... \right] \right\} , \qquad (2)$$

where $\bar{Q} = (2/3 - 1/3)/2 = 1/6$ is the average charge of the underlying fundamental fields (in this case the $u$ and $d$ quarks), $E_m$ is the end-point energy of the positron, $M$ is a hadronic mass of $\mathcal{O}$(1GeV), and the ellipsis stand for significant but smaller contributions that have been studied in detail. In the $^{14}O$ case, $E_m \approx 2.3$MeV and the first logarithmic term in Eq.(2) leads to a $\approx 3.45\%$ correction. This contribution literally rescues the SM from obvious contradiction ! A recent analysis[4] of the eight superallowed Fermi transitions leads to $V_{ud} = 0.9736 \pm 0.0007$. Combining this result with $V_{us} = 0.2205 \pm 0.0018$, and $V_{ub} = 0.004 \pm 0.002$, one obtains[4]

$$|V_{ud}|^2 + |V_{us}|^2 + |V_{ub}|^2 = 0.9965 \pm 0.0015 \ . \qquad (3)$$

This falls short of unity by 2.3 times the estimated error, which is mainly theoretical. Part of this uncertainty is due to the nuclear overlap correction $\delta_c$, for which there exist at present somewhat different competing evaluations. A more recent approach[4,5] attempts to take into account the contribution to $\delta_c$ from core nucleons, introducing a phenomenological correction factor $1 + aZ$ ($Z$ is the charge of the daughter nucleus), and determining $a$ from the data. This leads[4] to $V_{ud} = 0.9745 \pm 0.0007$ and

$$|V_{ud}|^2 + |V_{us}|^2 + |V_{ub}|^2 = 0.9983 \pm 0.0015 \ , \qquad (4)$$

which is consistent with unity. The agreement with universality is even better in Wilkinson's recent analysis (second paper of Ref.5), which is also based on a $Z$-dependent phenomenological correction factor ; he finds $V_{ud} = 0.97545 \pm 0.00082$ and $\sum_{i=d,s,b} |V_{ui}|^2 = 1.0001 \pm 0.0018$.[5] Thus, at present it is not clear whether there is disagreement or not but, if so, it is at the 0.3% rather than the 4% level, which would be devastating.

An interesting question is whether these are genuine electroweak corrections. To answer this query the following observations are relevant : i) One needs a renormalizable theory, such as the SM, to evaluate them. ii) It may be argued however, that the result can be reproduced with a local theory calculation involving only electromagnetic corrections, provided that one regularizes the result with a suitable cutoff $\Lambda$. iii) Point ii) can be answered by noting that the cutoff is important and that one needs the complete theory to determine it accurately. For example, if the cutoff were $\Lambda = v = (\sqrt{2}G_\mu)^{(-1/2)} = 246\text{GeV}$, a value which is rather reasonable and was in fact anticipated before the emergence of the SM, we would obtain

$$|V_{ud}|^2 + |V_{us}|^2 + |V_{ub}|^2 = \left\{ \begin{array}{l} 0.9919 \pm 0.0015 \\ 0.9937 \pm 0.0015 \end{array} \right. , \tag{5}$$

where the upper and lower entries correspond to the treatments leading to Eq.(3) and Eq.(4), respectively. These results differ from unity by $5.4\sigma$ and $4.2\sigma$, respectively, and are in clear disagreement with unitarity. Thus, an accurate determination of $\Lambda$ is necessary and this can only be provided by the complete theory. iv) One may also inquire what diagrams are relevant in the SM calculation. The analysis shows that one must consider the corrections associated with the complete gauge sector, not just the photon, and this includes all the vertex and box diagrams involving virtual $\gamma, Z^0$ and $W^\pm$. For example, the fermionic couplings of $\gamma$ and $Z^0$ are not universal, being different for leptons and quarks, and one must study all these diagrams in order to obtain meaningful results. Furthermore, it is only after all the gauge-sector contributions are combined that the amplitudes become convergent (after renormalization) and can be analyzed with short distance expansions, necessary for the control of strong-interaction effects ! On the other hand, the Higgs sector has an indirect effect: Eq.(2) holds exactly only when $\cos^2\theta_W = M_W^2/M_Z^2$ at the tree-level. We recall that this is the case in the presence of any number of Higgs doublets and singlets. When triplets and other representations are present so that $\cos^2\theta_W = M_W^2/\rho M_Z^2$, there is an additional contribution $[\alpha(\rho-1)/\pi]\ln(M_W^2/M_Z^2)[M_W^2/M_Z^2-1]^{-1}$ to the expression between curly brackets in Eq.(2).[2] However, this contribution is practically negligible, as current phenomenology shows that $\rho$ is very close to unity.

## 2. Evidence at High Energies.

I follow the discussion of Ref.6. An alternative approach is developed in Ref.7.

In the high-energy processes currently investigated the dominant electroweak corrections involve virtual fermions. Their effect is responsible for the large logarithms associated with the running of $\alpha$ [8] and the contributions from the $t-b$ isodoublet from which the $M_t$ constraints are derived.

It is natural to ask whether there is evidence in high-energy phenomena for corrections not contained in the running of $\alpha$ and, more specifically, in $\alpha(M_Z)$. One way to quantify this question is to "measure" $(\Delta r)_{res}$,[6] the residual part of $\Delta r$ [9] after extracting the effect of the running of $\alpha$. One has

$$\frac{\alpha}{1-\Delta r} = \frac{\alpha(M_Z)}{1-(\Delta r)_{res}} \ . \tag{6}$$

It is worth noting that $\alpha(M_Z)$ is scheme-dependent. Two frequently employed choices are : i) $\alpha(M_Z) = \alpha/(1-\Delta\alpha)$, where $\Delta\alpha = e^2 Re\left[\Pi^{(f)}_{\gamma\gamma}(0) - \Pi^{(f)}_{\gamma\gamma}(M_Z)\right]$ is the fermionic contribution to the conventional QED vacuum-polarization function. A recent determination gives $\Delta^{(5)}_{Rad} = 0.0280 \pm 0.0007$ for the five-flavour component which, when combined with the leptonic and very small top contributions, leads to $\alpha^{-1}(M_Z) = 128.899 \pm 0.090$.[10] Recent alternative evaluations are given in the papers of Ref.11. ii) The $\overline{\mathrm{MS}}$ definition $\hat{\alpha}(M_Z) = \alpha/[1 - e^2\Pi_{\gamma\gamma}(0)_{\overline{\mathrm{MS}}}]$, where the $\overline{\mathrm{MS}}$ subscript reminds us that the $\overline{\mathrm{MS}}$ renormalization has been implemented and $\mu = M_Z$ has been chosen. Updating the analysis of Ref.12 with the new result from Ref.10, one finds $e^2\Pi_{\gamma\gamma}(0)_{\overline{\mathrm{MS}}} = 0.0666 \pm 0.0007$ or $\hat{\alpha}^{-1}(M_Z) = 127.91 \pm 0.09$.

Inserting the direct world-average determination $M_W = 80.23 \pm 0.18$GeV [13] in the basic relation[9]

$$M_W^2(1 - M_W^2/M_Z^2) = (\pi\alpha/\sqrt{2}G_\mu)/(1 - \Delta r) \ , \tag{7}$$

one obtains $\Delta r = 0.0442 \pm 0.0104$. Using $\alpha^{-1}(M_Z) = 128.899$, Eq.(6) gives $(\Delta r)_{res} = -0.0161 \pm 0.0111$, which differs from zero by only $\approx 1.5\sigma$, not a very strong signal. The $M_W - M_Z$ interdependence, in conjuction with the experimental value of $M_W$, has also been extensively used by Z. Hioki[14] to examine the effect of various components of $\Delta r$.

The constraint is much sharper if we interpret the data in the framework of the fully fledged SM, treated as a Quantum Field Theory with its plethora of radiative corrections and interlocking relations.[6] The recent precision electroweak analysis, including all direct and indirect information, leads to $M_W = 80.32 \pm 0.06^{+0.01}_{-0.01}$ GeV,[15] where the last error reflects the uncertainty in $M_H$. Taking $M_W = 80.31 \pm 0.06$GeV, the worst case for the analysis, one finds $\Delta r = 0.0396 \pm 0.0035$ which implies $(\Delta r)_{res} = -0.0210 \pm 0.0037$. This value differs from zero by $5.6\sigma$. If we employ $\hat{\alpha}(M_Z)$ in Eq.(6) instead of $\alpha(M_Z)$, the evidence is even sharper : $(\Delta r)_{res}$ becomes $-0.0289 \pm 0.0037$, or $7.8\sigma$ away from a null result !

A scheme-independent argument can be obtained by considering two different definitions of $\sin^2\theta_W$ which are physical observables[6] : i) $\sin^2\theta^{lept}_{eff}$ ii) $\sin^2\theta_W = 1 - M_W^2/M_Z^2$. From the global fits one has : $\sin^2\theta^{lept}_{eff} = 0.2320 \pm 0.0003^{+0.0000}_{-0.0002}$ and $\sin^2\theta_W = 0.2242 \pm 0.0012^{+0.0003}_{-0.0002}$, and we see that they differ by $6.3\sigma$ ! As the two definitions agree at the tree level (because in the SM Lagrangian there is a single mixing angle), the difference must be due to radiative corrections. In particular, there is no B.A. involving a single mixing angle, whether related to $\alpha(M_Z)$ or not, that can accomodate all the information derived from the data using the full SM.

## 3. Evidence for Bosonic Electroweak Corrections in the SM.

I follow the discussion of Ref.16. There are also detailed studies based on an effective Lagrangian approach.[17]

By definition, at the one–loop level the electroweak bosonic corrections (E.B.C.)

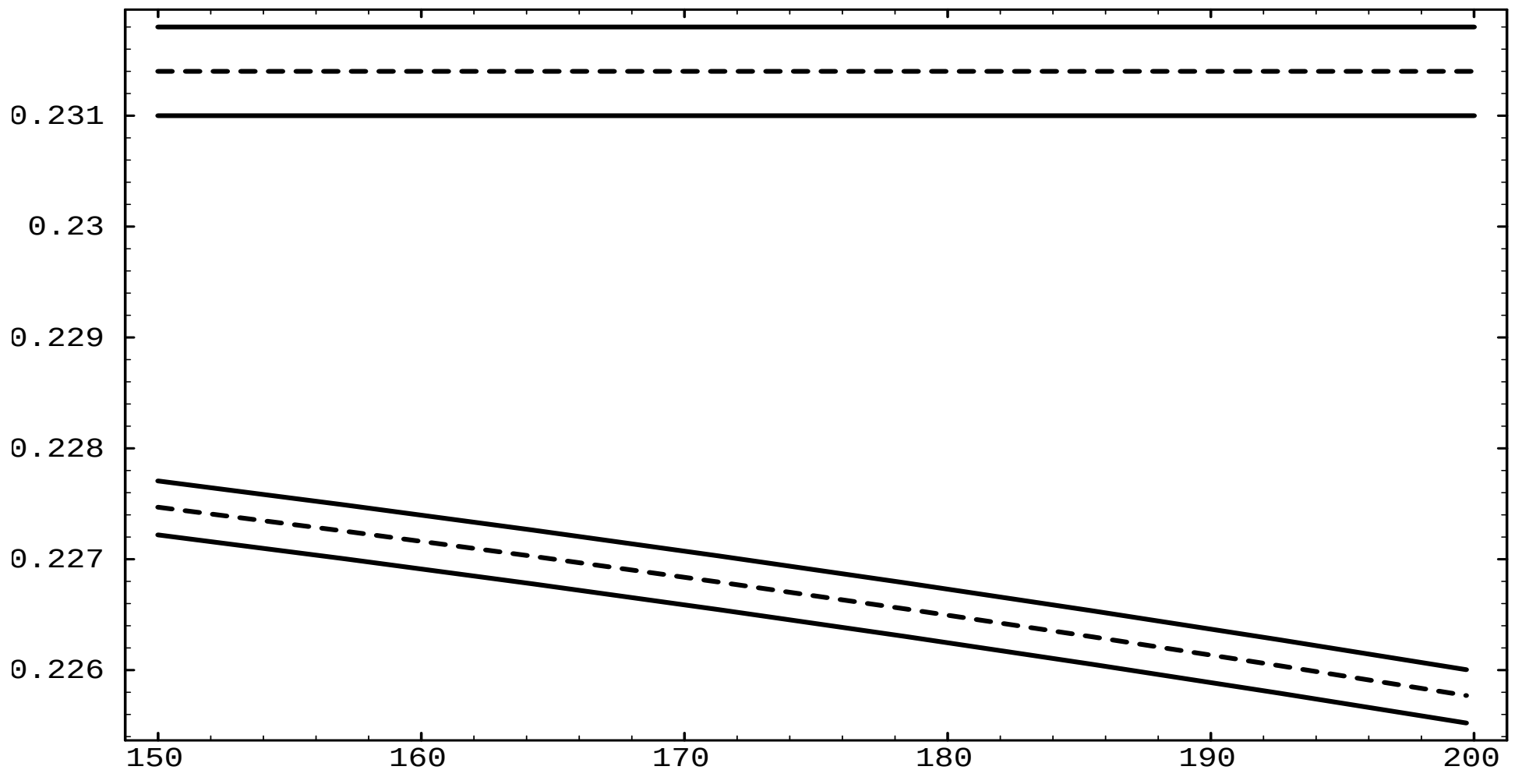

Fig. 1. Determination of $\sin^2\hat{\theta}_W(M_Z)$ from asymmetries (horizontal line) and ( $\alpha$, $G_\mu$, $M_Z$) (bottom curve) with the electroweak bosonic corrections removed, as a function of $M_t$(GeV).[14] The $1\sigma$ errors are indicated.

are all the contributions involving $W^\pm, Z^0, \gamma$, and $H$ as virtual particles in the loop. They include self-energy, vertex and box diagrams, and form a gauge-invariant sector (in the case of four fermion processes, the vertex and box diagrams also involve virtual fermions). In the processes currently explored these corrections are numerically subleading. However, they are very important conceptually, as they involve the fundamental particles and couplings of the gauge sector of the theory.

The current determinations of the weak-mixing angle are so precise that it is natural to inquire whether they are sensitive to the E.B.C.. The basic strategy outlined in Ref.16 is to determine the $\overline{\mathrm{MS}}$ parameter $\sin^2\hat{\theta}_W(M_Z)$ from two different sets of observables, subtracting the E.B.C. from the relevant radiative corrections, and finding out whether the two results are consistent (see Fig.1). Specifically, one determines $\sin^2\hat{\theta}_W(M_Z)$ from the assymetries via the relation[18]

$$\sin^2\theta_{eff}^{lept} = Re\hat{k}_\ell({M_Z}^2)\hat{s}^2 \ , \tag{8}$$

where $\hat{s}^2$ is an abbreviation for $\sin^2\hat{\theta}_W(M_Z)$ and $\hat{k}_\ell(M_Z)$ is the electroweak form-factor multiplying $\hat{s}^2$ in the $Z^0 \rightarrow \ell\bar{\ell}$ amplitude. The second determination is from $G_\mu$, $\alpha$, and $M_Z$, and can be implemented from the basic relation[19]

$$\hat{s}^2\hat{c}^2 = \frac{\pi\alpha}{\sqrt{2}G_\mu {M_Z}^2(1-\Delta\hat{r})} \ , \tag{9}$$

where $\Delta\hat{r}$ is the radiative correction. In fact, it has been noted that the E.B.C. to $\Delta\hat{r}$ are quite sizeable, namely $(\Delta\hat{r})_{E.B.C.} = 0.97\%,\ 1.22\%,\ 1.57\%$ for $M_H = 10,\ 100,\ 1000$

GeV, respectively.[19] By way of comparison $(\Delta r)_{E.B.C.} = -0.22\%,\ 0.30\%,\ 1.18\%$ for $\xi \equiv M_H^2/M_Z^2 = 0,\ 1,\ 100$.[9] A recent analysis[18] shows that $\hat{k}_\ell(M_Z{}^2) = 1.0012 \pm i0.0134$. The very small deviation of $Re\hat{k}_\ell(M_Z{}^2)$ from unity is due to a fortuitous cancellation of larger radiative corrections of $\mathcal{O}(\hat{\alpha}/2\pi\hat{s}^2 \approx 0.5\%)$. If the E.B.C. corrections are removed, one finds instead $Re\hat{k}_\ell(M_Z{}^2)_{tr} = 1.0060$, where the $tr$ subscript reminds us that the calculation has now been performed on the basis of a truncated version of the theory. Employing the value $\sin^2\theta_{eff}^{lept} = 0.2317 \pm 0.0004$ obtained from the asymmetry measurements at LEP and SLC,[15] Eq.(8) gives

$$(\sin^2\hat{\theta}_W)_{tr} = 0.2303 \pm 0.0004 \quad \text{(asymmetries)}\ . \tag{10}$$

The figure, an updated version of Fig.1 of Ref.16, compares Eq.(10) with the values extracted from $(\hat{s}^2)_{tr}(\hat{c}^2)_{tr} = (\pi\alpha/\sqrt{2}G_\mu M_Z{}^2)/(1-(\Delta\hat{r})_{tr})$, as a function of $M_t$ (note that $(\Delta\hat{r})_{tr}$ is independent of $M_H$).

It is apparent that the removal of the E.B.C. leads to a sharp disagreement. At the lower bound $M_t = 150$GeV of Fig.1, the value extracted from $\alpha$, $G_\mu$, $M_Z$, is $(\hat{s}^2)_{tr} = 0.2275 \pm 0.0003$, which differs from Eq.(10) by $5.6\sigma$. For $M_t = 180$GeV, the discrepancy reaches $7.6\sigma$ ! The removal of the E.B.C. leads to lower $\hat{s}^2$ values in both determinations. The inconsistency arises because the effect is much more pronounced in the $(\alpha,\ G_\mu,\ M_Z)$ analysis.

It is natural to inquire whether one can find signals for the $H$ boson contribution by removing it from the corrections, retaining the rest. As $H$ does not contribute to $\hat{k}_\ell(M_Z{}^2)$ at the one-loop level, its removal does not affect the SM determination of $\hat{s}^2$ from the asymmetries, which is $\hat{s}^2 = 0.2314 \pm 0.0004$. The removal of the Higgs contribution from $\Delta\hat{r}$ must be done in a finite and gauge invariant manner. In the SM the sum of the diagrams involving $H$ in the self-energies contributing to $\Delta\hat{r}$ is gauge invariant, but divergent. Therefore, one must specify the renormalization prescription and the scale at which they are evaluated. A natural possibility is to subtract the $\overline{\text{MS}}$-renormalized $H$ contribution evaluated at $\mu = M_Z$. In this case, one has[16]

$$(\Delta\hat{r})_{HB} = \frac{\alpha}{4\pi\hat{s}^2}\left[\frac{1}{\hat{c}^2}H(\xi) - \frac{3}{4}\frac{(\xi\ln\xi - \hat{c}^2\ln\hat{c}^2)}{\xi - \hat{c}^2} + \frac{19}{24} + \frac{\hat{s}^2}{6\hat{c}^2}\right]\ , \tag{11}$$

where $\xi \equiv M_H^2/M_Z{}^2$ and $H(\xi)$ is a function given in Ref.9. In contrast with the situation when the full E.B.C. are removed, the analysis shows that the subtraction of $(\Delta\hat{r})_{HB}$ does not lead to an incosistent result. This is easy to understand, as $(\Delta\hat{r})_{HB}$ vanishes for $M_H \approx 113$GeV. Thus, the subtraction of $(\Delta\hat{r})_{HB}$ is equivalent to a SM calculation with $M_H \approx 113$GeV, which is consistent with the electroweak data.

In summary, we have presented strong indirect evidence for the presence of E.B.C. in the SM. If one probes just the Higgs component in the particular way we have outlined, no evidence has been uncovered in this very simple analysis.

## 4. Acknowledgment

The author would like to thank P. Gambino for useful discussions and for providing an updated version of Fig.1. This work was supported in part by the National Science Foundation under Grant No.PHY-9313781.